\documentclass[twocolumn,superscriptaddress,showpacs,floatfix,nofootinbib,nobalancelastpage,raggedbottom,aps,pre,10pt,amssymb,amsfonts,amsmath,eqsecnum]{revtex4-1}
\usepackage[utf8]{inputenc}
\usepackage{ amsbsy,   appendix, bm, booktabs, csquotes, dcolumn, dsfont, graphicx, multirow, longtable, mathtools, matlab-prettifier}

\definecolor{col1}{RGB}{39,118,71}
\definecolor{col2}{RGB}{180,204,78}
\definecolor{col3}{RGB}{255,198,0}
\definecolor{col4}{RGB}{133,58,118} 
\definecolor{col5}{RGB}{75,12,59}
\definecolor{col6}{RGB}{0,0,0}
\usepackage[colorlinks=true,
            linkcolor=col1,
            urlcolor=col4,
            citecolor=col2,
            anchorcolor=col6]{hyperref}
\usepackage[T1]{fontenc}
\usepackage[margin=1in]{geometry}
\setcitestyle{square}
\PassOptionsToPackage{linktocpage}{hyperref}

\newcommand{\bs}[1]{{\boldsymbol{#1}}}
\newcommand{\Cdot}{\! \cdot \!}
\newcommand{\nhat}{\hat{\bs{n}}}

\mathchardef\mhyphen="2D
\bibstyle{aps}

\begin{document}
\title{{Configurational Information Measures, Phase Transitions,\\ and an Upper Bound on Complexity}}
\author{Damian R Sowinski}
\email{Damian.Sowinski@rochester.edu}
\author{Sean Kelty}
\author{Gourab Ghoshal}
\affiliation{Department of Physics and Astronomy, University of Rochester}

\date{\today}
\begin{abstract}
Configurational entropy (CE) and configurational complexity (CC) are recently popularized information theoretic measures used to study the stability of solitons. 
This paper examines their behavior for 2D and 3D lattice Ising Models, where the quasi-stability of fluctuating domains is controlled by proximity to the critical temperature.
Scaling analysis lends support to an unproven conjecture that these configurational information measures (CIMs) can detect (in)stability in field theories.
The primary results herein are the derivation of a model dependent CC-CE relationship, as well as a model independent upper bound on CC.
CIM phenomenology in the Ising universality class reveals multiple avenues for future research.
\end{abstract}
\maketitle
\section{Introduction}
Configurational Information Measures (CIMs) were introduced by Gleiser and Stamatopolous (GS12) as a framework for quantifying structural complexity in nonlinear field theories, particularly in the context of solitonic solutions in $(1\!+\!1)$-dimensional scalar field models~\cite{gleiser2012entropic}. 
Rooted in Shannon entropy in momentum space~\cite{Shannon1948communication}, GS12 proposed the notion of configurational entropy (CE), demonstrating that it minimizes for field configurations approximating exact solitonic solutions. 
This initial formulation led to a surge of investigations exploring CE across a broad spectrum of physical systems, ranging from high-energy physics~\cite{correa2016bounds,bernardini2017configurational,gleiser2018bconfigurational,braga2018thermal,ma2018shannon,gleiser2019resonant,bazeia2019configurational,ferreira2019pion,thakur2020configurational,braga2020configuration,thakur2021configurational}, to astrophysics and cosmology~\cite{gleiser2013information,correa2015configurational,gleiser2015stability,casadio2016stability,correa2016configurational,braga2017configurational,alexander2018entropy,wibisono2018information,braga2019information,fernandes2019extended,bernardini2019cosmological,da2021ads,da2021information,barreto2022differential}, and statistical physics, including phase transitions and critical phenomena~\cite{gleiser2012cinformation,gleiser2014transition,Sowinski2015information,correa2015entropic,sowinski2016complexity,sowinski2017information,barbosa2018dynamical,lee2019configurational,lima2021phase}.

As these studies progressed, refinements in CE's definition and computation gave rise to a broader class of measures, collectively termed CIMs~\cite{gleiser2018configurational}. 
A key emergent feature of CIMs is their apparent ability to identify stability conditions in a diverse range of dynamical systems. 
Specifically, CIM extrema have been empirically linked to critical behavior in thermal field theories~\cite{Sowinski2015information,sowinski2017information}, soliton stability and decay\cite{gleiser2018bconfigurational}, and equilibrium conditions in self-gravitating systems such as polytropic white dwarfs and Tolman-Oppenheimer-Volkoff neutron stars~\cite{gleiser2012cinformation,gleiser2015stability}. 
These observations suggest that the spatial structure of a system encodes information about its temporal stability, where extrema in CIMs correspond to transitions between stable and unstable equilibria.
Phenomenological support has steadily increased for this {\it stability conjecture} (SC) in the past two decades, no analytical tools, to the authors' best knowledge, have been developed to better understand why it should be the case.

The primary contribution of this paper is the derivation of a model dependent analytical relationship between CIMs, and, as a corollary, a model independent upper bound on the CC.
The secondary contribution is to extend the known phenomenological behavior of CIMs to the paradigmatic Ising model~\cite{goldenfeld_lectures}, both in two and three spatial dimensions, in further support of the SC. 
Previous studies demonstrated that CE minimizes at critical points in the Ginzburg-Landau mean field model (GL)~\cite{Sowinski2015information,sowinski2017information}, where diverging correlation times prolong fluctuation lifetimes. 
The Ising model, to whose universality class GL belongs to, provides a well-controlled setting for analyzing how information-theoretic measures track stability across different scales. 
Understanding CIM behavior in this model not only strengthens the connection between information theory and statistical physics but also lays an outline for applications in other universality classes.

In Section~\ref{sec: II}, we provide an overview of the theoretical development of CIMs, focusing on foundational work in~\cite{Sowinski2015information,sowinski2016complexity,sowinski2017information} and its role in formulating the SC. 
We adopt the modern terminology established in~\cite{gleiser2018configurational,gleiser2018bconfigurational},which distinguishes CE from CC, a clarification crucial for readers engaging with earlier literature. 
To contextualize our study, we also provide a concise history of the Ising model, adding to the growing body of work analyzing CIM phenomenology across diverse models. 
Section~\ref{sec: III} outlines the details of our numerical implementation, including the methodology for sampling Ising states and estimating CIMs.
In Section~\ref{sec: IV}, we present and interpret the observed behavior of CIMs, culminating in the derivation of the promised CIM relationship and the upper bound on CC. 
Finally, in Section~\ref{sec: V}, we discuss the implications of our results. 

\section{Background}\label{sec: II}

Despite being a relatively recent development, introduced only fifteen years ago, the terminology and definitions surrounding CIMs have evolved significantly. 
In this section, we provide an overview of their historical development, using the most up-to-date formalism as of 2025. 
We exclude discussion of differential CIMs, directing interested readers to Ref.~\cite{sowinski2017information}. 
Additionally, we present a brief historical account of the Ising model, which serves as the primary model investigated in this work.

\subsection*{Configurational Information Measures}

The notion of CC first emerged in the study of ranking different approximations to scalar field solitons in classical field theory~\cite{gleiser2012cinformation}. 
Typically, trial functions, $\phi(x)$, are used to approximate solitonic solutions, with optimization performed by minimizing their energy. 
However, when multiple ans\"atze yield degenerate energy values, a complementary measure is required to distinguish between them. 
This challenge motivated the introduction of CE, constructed analogously to Shannon entropy but defined in momentum space. 
Using the normalized power spectrum of the ans\"atze,
\begin{equation}\label{eq: normalized PS}
    \mathcal P(k) =\frac{1}{Z}|\tilde \phi(k)|^2, \quad \text{where} \quad Z=\sum_k|\tilde\phi(k)|^2, 
\end{equation}
with $\tilde\phi$ the Fourier transform of the trial function, CE is defined as
\begin{align}\label{eq: CE1}
    CE = -\sum_k \mathcal{P}(k)\log\mathcal{P}(k),
\end{align}
providing a measure of how localized the field distribution is in momentum space. 

To facilitate extension to continuous fields, the modal fraction was introduced by normalizing the power spectrum by its maximum value:
\begin{align}\label{eq: modal fraction}
    f(k) = \frac{\mathcal P(k)}{\max_{k'}\mathcal P(k')}.
\end{align}
This refinement altered the measure fundamentally, transitioning it from an entropic measure to one of complexity,
\begin{align}\label{eq: CC1}
    \text{CC}=-\sum_k f(k)\log f(k),
\end{align}
though this was not appreciated at the time.
Analysis across multiple solitonic structures demonstrated that CC generally attains higher values in approximate ans\"atze than in true solutions, reinforcing its role as a discriminator between stable and metastable configurations.
Unfortunately, the precise construction of the CC remained nebulous; for analyzing kinks the energy density was used to construct the modal fraction, while for the analysis of bubbles it was the field itself.

To address cases where CC unexpectedly underestimated complexity, an extremum principle was introduced through the concept of Relative Configurational Entropy (RCE):
\begin{align}
    RCE = \frac{|\text{CC}_\text{ansatz} - \text{CC}_\text{true}|}{\text{CC}_\text{true}}.
\end{align}
By construction RCE successfully identified optimal approximations, its dependence on knowledge of the exact solution, however, limited its applicability. 
To circumvent this, a generalization in terms of Kullback-Leibler divergence (KLd) was proposed, defining RCE as the divergence between the power spectra of two states:
\begin{align}
    RCE &= D_{KL}(\mathcal{P}||\mathcal{P}_0)\nonumber\\
    &=\sum_k \mathcal{P}(k)\log \frac{\mathcal{P}(k)}{\mathcal{P}_0(k)},
\end{align}
where $\mathcal{P}_0(k)$ corresponds to a reference power spectrum, often taken from thermal equilibrium approximations~\cite{Aarts2000exact}.
Empirical studies found RCE correlates strongly with the number of oscillons present in post-quench scalar field dynamics ~\cite{gleiser2013information,gleiser2014transition}. 
Unlike thermodynamic entropy, which maximizes at equilibrium, RCE attains peak values far from equilibrium, implying its interpretation as a measure of spatio-temporal complexity.

In \cite{gleiser2013information} CC was applied to non-topological solitons of complex scalar fields known as Q-balls. 
A classical stability criterion identifies when large Q-balls destabilize into diffuse Q-clouds; the maximal value of CC was shown to correlate very closely with the criterion, pointing towards a relationship between CC and stability. 
A similar study was performed on polytropic models of stars, with a small change in the definition needed to describe objects with a finite radius:
The modal fraction was defined only for modes whose wavelengths were smaller than the diameter of the object, arguing that there is no informational contribution coming from length-scales larger than those needed to describe an object.
Remarkably, this modified CC, divided by the central density of the polytrope, maximized at the well known Chandrasaekhar limit, and minimized in the limit of a cold, stable white dwarf.
This trend extended to astrophysical systems; polytropic stellar models displayed maxima in CC at the Chandrasekhar limit, beyond which gravitational collapse ensues. 
Similar findings emerged in the analysis of neutron stars, where CC extrema delineated regions of perturbative stability and relativistic instability~\cite{gleiser2015stability,Koliogiannis2023}.

Parallel efforts investigated the role of CE in critical phenomena. 
In Ginzburg-Landau models, CE exhibited marked reductions at phase transitions, reflecting changes in the underlying correlation structure of fluctuations~\cite{gleiser2015information, sowinski2017information}. 
Such behavior is expected, as the power spectrum narrows near criticality due to an increasing dominance of long-wavelength fluctuations. 
Importantly, this work was the first to suggest examining the CE density through the direction-averaged information storage,
\begin{align}
    ce(k) &=\frac{1}{\Omega_d}\int d^d\nhat\  \mathcal{P}(k\nhat) I( k\nhat),
\end{align}
where $I(\bs{k})=\log\mathcal P(\bs{k})$ is the pointwise information, and $\Omega_d$ the $d$-dimensional solid angle.
The examinations identified three distinct regimes of modes storing configurational information---scale-free, turbulent, and critical---each characterized by unique spectral properties~\footnote{It should be noted that there is a mistake in  equation 12 of \cite{sowinski2017information}. It is valid in $d=2$ dimensions, but not so for $d\neq 2$. Fortunately, this does not affect the paper's other results.}. 
Since, at the time, there was no clear delineation between CE and CC, the studies did not look at latter.

While previous investigations established CIM phenomenology in multiple physical settings, a comprehensive study of both CC and CE within the Ising universality class remains absent. 
The present work fills this gap by systematically analyzing both, providing further insight into the fundamental links between configurational measures and stability properties.

\subsection*{Ising Model}

The Ising model is a fundamental system in statistical physics, providing a minimal lattice representation of ferromagnetism~\cite{weiss1907hypothesis,lenz1920beitrag}.
Classically, it consists of a $d$-dimensional cubic lattice with $N = L^d$ sites, each occupied by a spin variable $\sigma_i \in \{-1, +1\}$. The Hamiltonian governing the system is given by:
\begin{align}\label{eq: Ising Hamiltonian}
    H[\bs{\sigma}]= -\frac{J}{2} \bs{\sigma}^T \Cdot \bs{A} \Cdot \bs{\sigma},
\end{align}
where $J > 0$ is the ferromagnetic coupling constant, and $\bs{A}$ is the adjacency matrix defining nearest-neighbor interactions. The system is coupled to a heat bath at temperature $T$, with equilibrium configurations sampled from the Boltzmann distribution.

A key quantity characterizing local interactions is the energy of each spin, defined as:
\begin{equation}\label{eq: energy density}
    \bs{\varepsilon}=-\frac{J}{2}\bs{\sigma}^T\Cdot \bs{A}\odot\bs{\sigma},
\end{equation}
where $\odot$ is pointwise multiplication.
This quantity provides insight into the spatial organization of domains by tracking their boundaries:
since $H=\mathds{1}^T\bs{\varepsilon}$, the Hamiltonian ``counts'' the surface area of all the domains.

\begin{figure}[ht]
    \centering
    \includegraphics[width=0.9\linewidth]{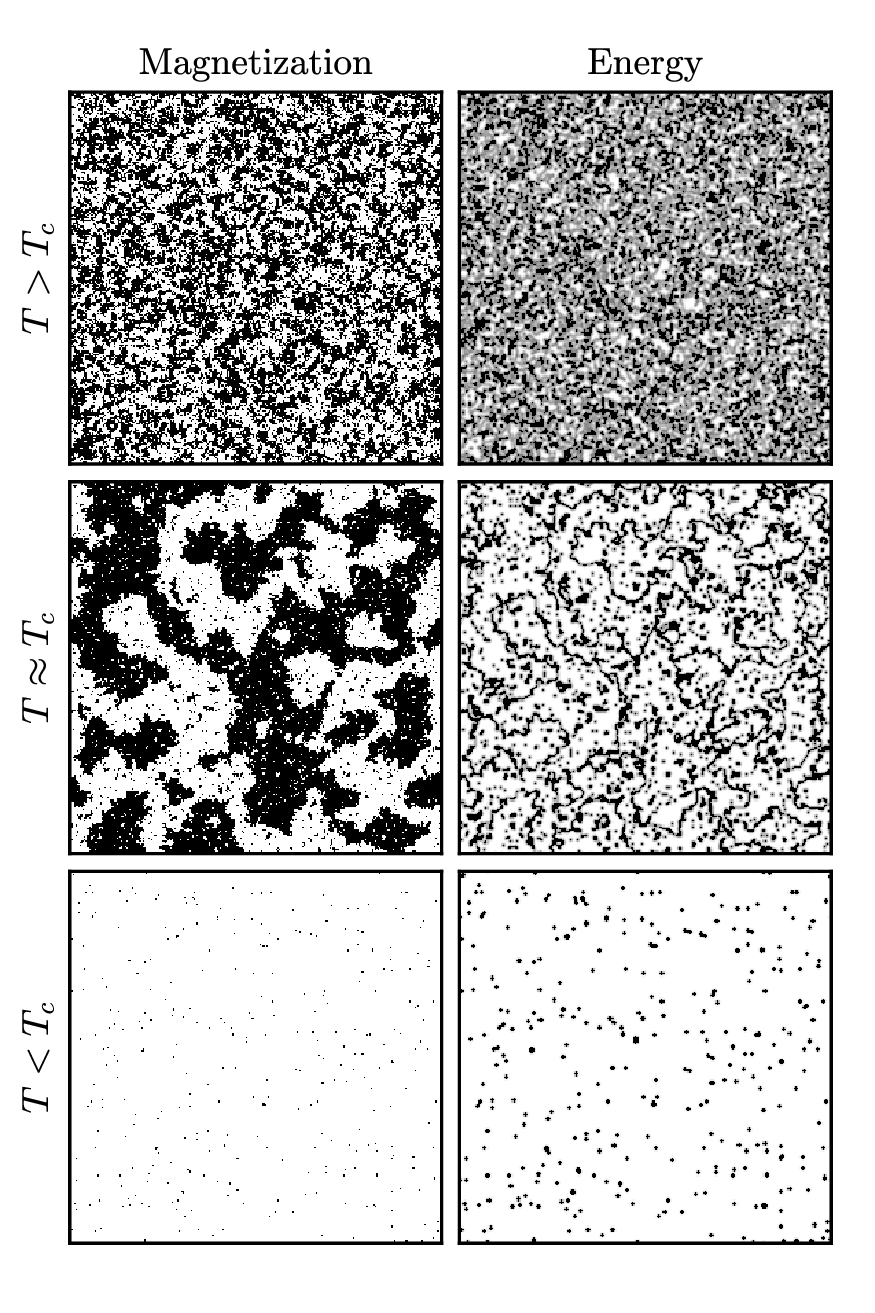}
    \caption{Representative configurations of the 2D Ising model at different temperatures. The left column shows spin configurations, while the right column displays the corresponding energy distributions based on Eq.~\eqref{eq: energy density}. The middle row corresponds to a temperature near the critical point $T_c$, with the top and bottom rows depicting states at higher and lower temperatures, respectively.}
    \label{fig: ising states}
\end{figure}

The system undergoes a second-order phase transition at a critical temperature $T_c$, where the magnetization, $m = \langle \sum_i \sigma_i \rangle / N$, serves as an order parameter.
Below $T_c$, spontaneous symmetry breaking occurs, leading to nonzero magnetization, while above $T_c$, the system remains disordered. Figure~\ref{fig: ising states} illustrates spin and energy configurations at various temperatures.

The 1D Ising model, solved by Ernst Ising in 1924~\cite{ising1924beitrag}, lacks a phase transition at finite temperature. However, Peierls' argument~\cite{peierls1936ising} and subsequent analytical results confirmed the existence of a finite $T_c$ in two and higher dimensions. The exact solution for the 2D model, provided by Onsager~\cite{onsager1944crystal}, determined critical exponents and established the Ising universality class.

Phase transitions in the Ising model are linked to diverging correlation lengths and times, making it a paradigmatic system for studying not only critical phenomena, but the stability of fluctuations. 
The effective Ginzburg-Landau Hamiltonian provides a coarse-grained description of this transition:
\begin{align}\label{eq: GL Hamiltonian}
    H_{GL}\! =\! \frac{1}{2}\int_{\mathds{R}^d}\!\!\! \!\!d^d\varphi\left(\!\dot\varphi^2 + |\nabla\varphi|^2 + (\frac{T}{T_c}\!-\!1)\varphi^2 + \frac{1}{2}\varphi^4\!\right).
\end{align}
Through the Hubbard-Stratonovich transformation~\cite{stratonovich1957method}, the Ising model maps directly to the non-kinetic terms of Eq.~\eqref{eq: GL Hamiltonian}, reinforcing their shared universality class. Since Configurational Entropy (CE) has been previously studied in the context of Ginzburg-Landau models, its application to the Ising model provides a natural extension for exploring the relationship between CIMs and phase transitions.

\begin{figure*}[t!]
    \centering
    \includegraphics[width=0.49\textwidth]{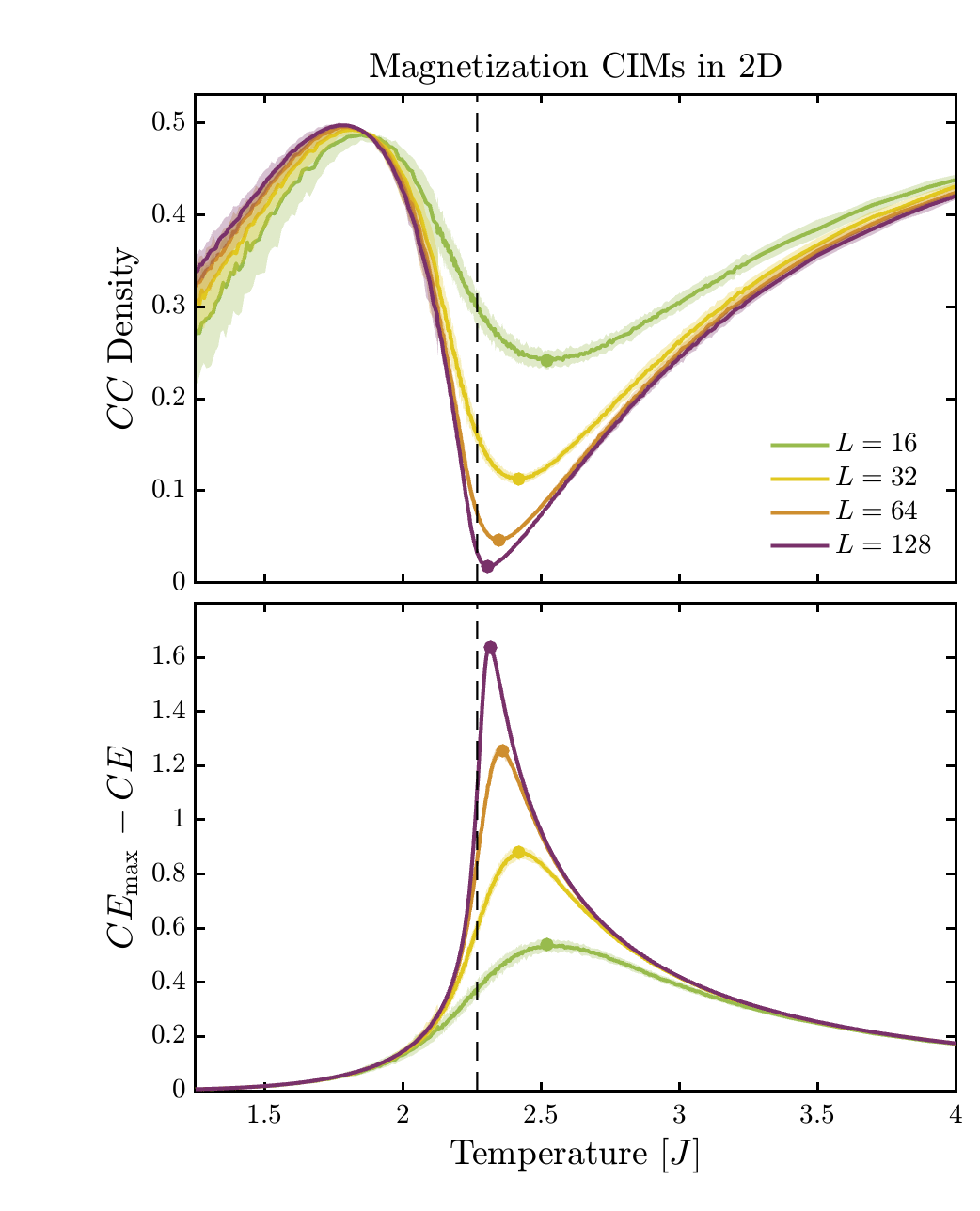}
    \includegraphics[width=0.49\textwidth]{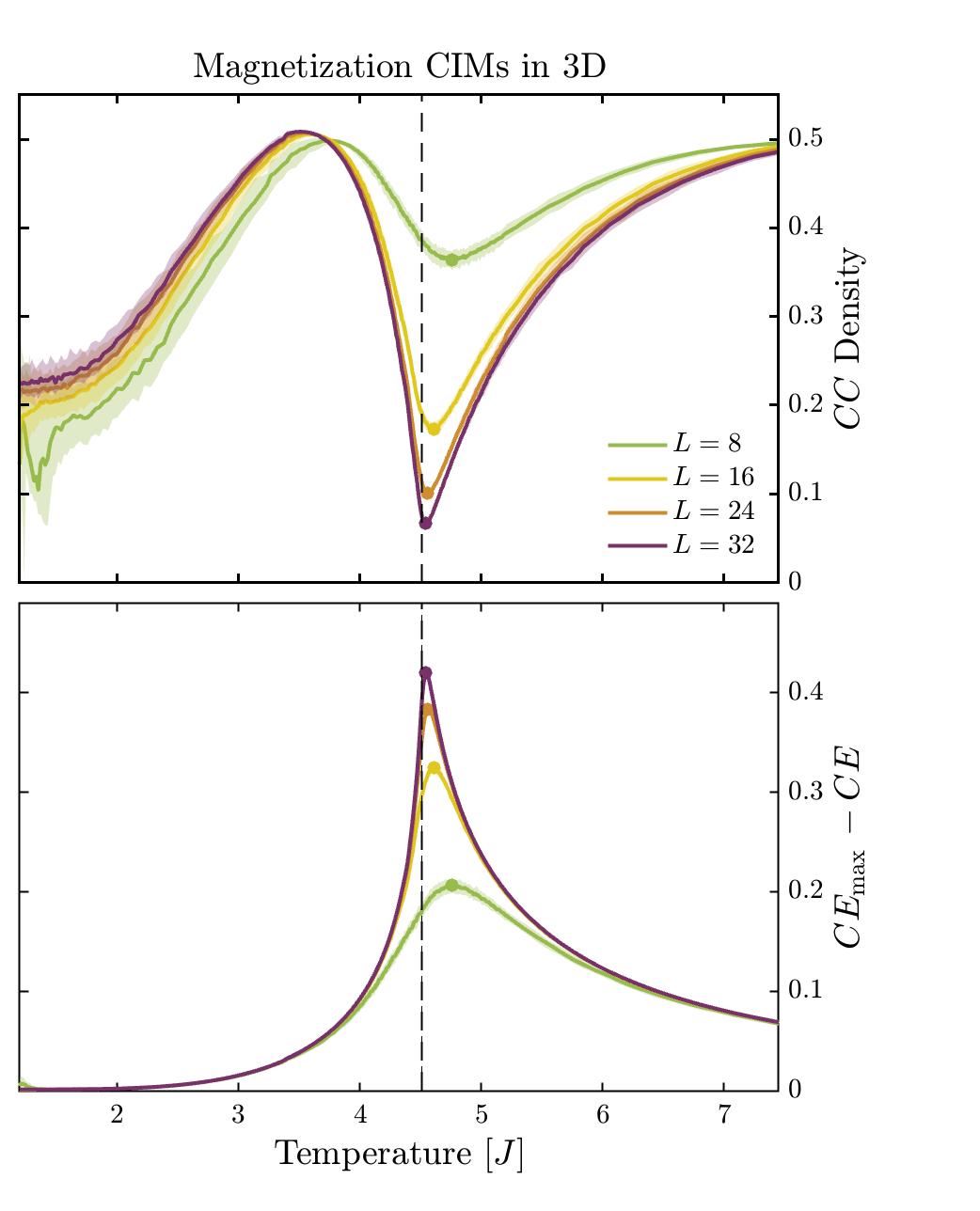} 
    \caption{Temperature dependence of configurational information measures (CIMs) for magnetization in the Ising model. The left and right panels correspond to the 2D square and 3D cubic lattices, respectively. The top row displays the configurational complexity per spin, while the bottom row shows the deviation of configurational entropy from its maximal value, $\log L^d$. The shaded regions represent the 80\% confidence interval from ensemble fluctuations, with minima marked by circles. The vertical dashed lines indicate the thermodynamic critical temperatures.}
    \label{fig: CIMs magnetization}
\end{figure*}

\section{Methods}\label{sec: III}

To analyze the behavior of Configurational Information Measures (CIMs) in the Ising model, we generate data ensembles using custom parallelized MATLAB code on the Bluehive Cluster at the Center for Integrated Research Computing at the University of Rochester~\cite{bluehive}. 
Sampling is performed for both the 2D and 3D Ising models on periodic square and cubic lattices, respectively, across various system sizes: $L=16,32,64,128$ in 2D and $L=8,16,24,32$ in 3D. 
For each lattice size, $100$ independent realizations are simulated across $313$ values of the inverse temperature $\beta$ in 2D, and $230$ values in 3D.

The inverse temperature ranges are chosen to encompass the known critical temperatures: $\beta \in [0.30772,0.81072]$ for 2D and $\beta \in [0.132,1.009]$ for 3D, ensuring coverage around $\beta_c=2/\ln(1+\sqrt{2})\approx0.4407$ for 2D and $\beta_c\approx0.221654626(5)$ for 3D~\cite{Onsager, 3D_critical_temp}. 
Near criticality, fine sampling is conducted with $\Delta\beta = 5 \times 10^{-4}$, while further from criticality, a coarser grid is used with steps increased by factors of $6$, $10$, or $20$ depending on the temperature range. 
Results are presented in terms of the dimensionless temperature $T = k_B / (\beta J)$.

Each simulation follows a two-stage heating and sampling scheme. The lattice is initialized in an ordered state ($\forall i, \sigma_i=1$) and equilibrated at the lowest sampled temperature using Metropolis-Hastings (MH) updates~\cite{metropolis1953equation} for $2000$ steps. 
This is followed by a sampling phase of $2000$ steps, with a configuration recorded every fifth step. 
The temperature is then incrementally increased, with shorter re-equilibration ($500$ steps) followed by another sampling period. 

In the vicinity of the critical temperature ($\beta = 0.47002$ for 2D and $\beta = 0.294$ for 3D), a hybrid sampling approach is employed. 
Heating remains the same, but sampling transitions to the Wolff algorithm~\cite{wolff1989collective}, where clusters are flipped until $2N$ spin flips have occurred before recording a configuration. 
This process repeats for $250$ cycles before advancing to the next temperature. 
This hybrid method is used for all subsequent temperatures.

At each temperature, data from independent realizations are pooled into ensembles. 
In 2D, ensembles contain $4\times 10^4$ samples for each of the $91$ MH sampled temperatures, and $2.5\times 10^4$ samples for the $222$ Wolff sampled ones. 
The 3D ensembles follow the same structure but with $74$ MH and $156$ hybrid temperatures.

Data extraction follows a structured pooling scheme at fixed temperature and lattice size, enabling statistical quantification of our CIM estimators. 
Each sample is labeled by a Greek index $\alpha$ and partitioned by the originating lattice realization, indexed by capital Latin letters, where $|A|$ denotes the number of samples in partition $A$. 
The mean magnetization and energy per spin for each sample are computed as:
\begin{align}
    \bar{\sigma}^\alpha &= \frac{1}{N} \sum_{i=1}^N \sigma_i, \quad \bar{\varepsilon}^\alpha = \frac{H}{N},
\end{align}
allowing calculation of fluctuation fields:
\begin{align}
    \delta \bs{\sigma}^\alpha &= \bs{\sigma}^\alpha - \bar{\sigma}^\alpha, \quad \delta \bs{\varepsilon}^\alpha = \bs{\varepsilon}^\alpha - \bar{\varepsilon}^\alpha.
\end{align}
Power spectra are estimated by first applying a discrete Fourier transform (DFT) to these fluctuation fields, taking the squared modulus pointwise, and finally averaging over all the samples belonging to the same lattice realization,
\begin{align}
    \hat{\bs{\mathcal P}}^A = \frac{1}{|A|} \sum_{\alpha \in A} \texttt{fftn}(\delta \bs{\sigma}^\alpha)^* \odot \texttt{fftn}(\delta \bs{\sigma}^\alpha).
\end{align}
DFTs are computed using MATLAB's built-in \texttt{fftn} function~\cite{MATLAB}.

Power spectra are then processed to compute CIMs, the only difference being how power spectra are normalized. 
For CE the spectra are divided by their $L_1$-norm, while for CC they are divided by their maximum values.
Reported CIM values correspond to the unbiased estimator of the mean over partitions, with uncertainties given by the unbiased estimator of the standard deviation.

\begin{figure*}[t!]
    \centering
    \includegraphics[width=0.49\textwidth]{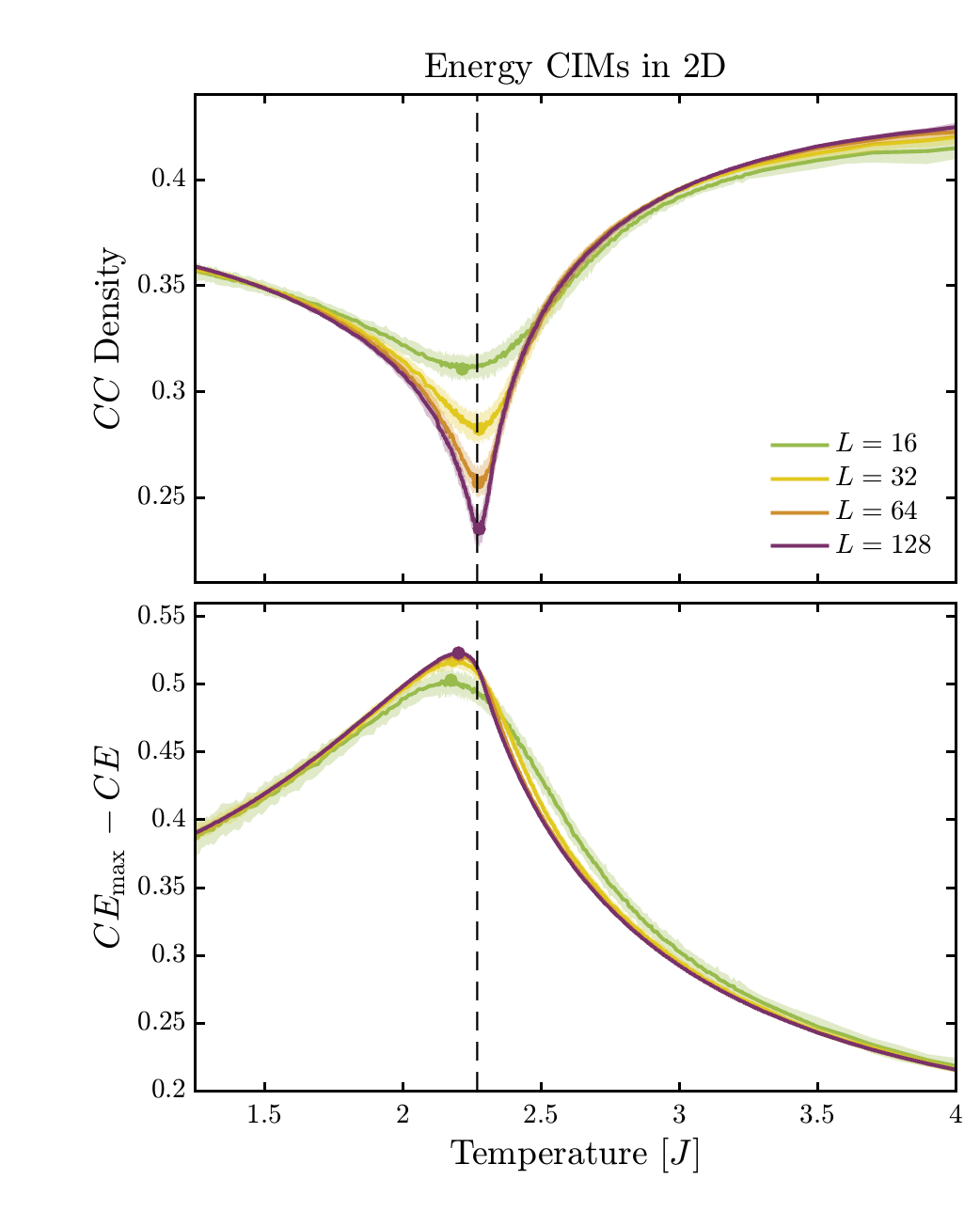}
    \includegraphics[width=0.49\textwidth]{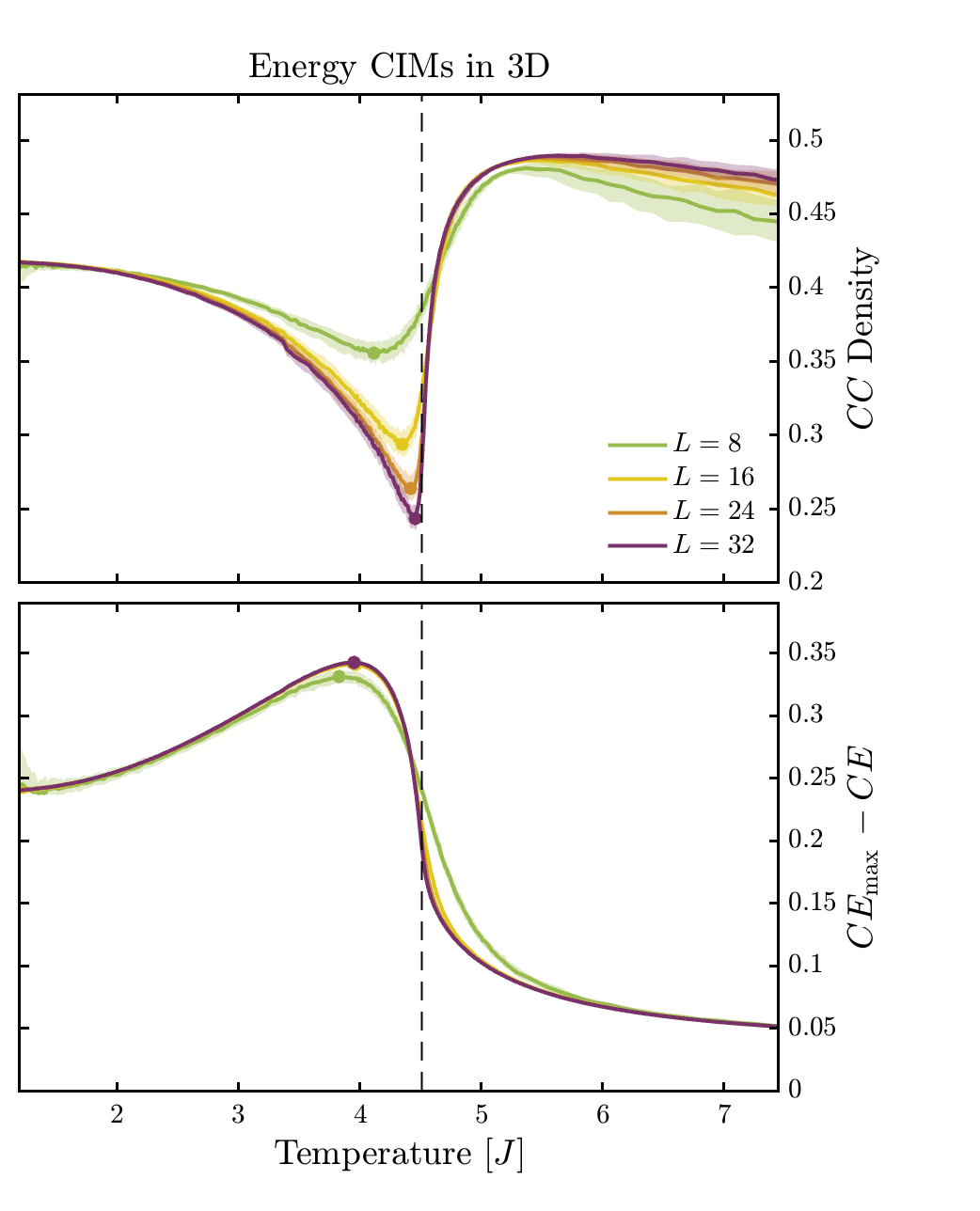} 
    \caption{Temperature dependence of configurational information measures (CIMs) for energy fluctuations in the Ising model. The left and right panels correspond to the 2D square and 3D cubic lattices, respectively. The layout and descriptions follow Fig.~\ref{fig: CIMs magnetization}, demonstrating consistency between magnetization- and energy-based CIMs.} 
    \label{fig: CIMs energy}
\end{figure*}

\section{Results}\label{sec: IV}

We now present the behavior of configurational entropy (CE) and configurational complexity (CC) across the range of investigated temperatures, linking these findings to our previous theoretical framework and numerical methodologies. 
Figures~\ref{fig: CIMs magnetization} and~\ref{fig: CIMs energy} display simulation results for both 2D and 3D Ising models, focusing on magnetization and energy configurations, respectively.

CE is expected to be near its maximal value in both high and low temperature limits. 
In the high-temperature regime, $T \to \infty$, thermal noise dominates, leading to a nearly flat power spectrum. 
In the low-temperature regime, long-wavelength modes dominate, but the fluctuation power spectrum, which excludes background magnetization, remains approximately white.
For a white spectrum CE$_{\max} =\log L^d$ grows with $L$.
Dividing the CE by CE$_{\max}$ turns out to not give as clean of a scaling analysis as one might expect; surprisingly, the difference from it does.
The drop from maximum CE is plotted in the bottom panels of both figures~\ref{fig: CIMs magnetization} and~\ref{fig: CIMs energy}, behavior aligning with the Ginzburg-Landau model's CE phenomenology~\cite{Sowinski2015information,sowinski2017information}.
The data does not preclude the CE drops asymtptoting to non-zero values in either limit, except in the case of the magnetization CE at low temperatures.
Secondly, for each lattice size in either dimension, the drop in magnetization CE maximizes above the critical temperature, and below it for the energy CE.
This mimics the well-known finite size scaling behavior of the susceptibility and specific heat, which too are built from the fluctuation fields of the spin and energy, respectively~\cite{newman_barkema}.
However, the two physical quantities use powers of the temperature in their definitions, which suppresses the high temperature behavior of the fluctuating field statistics.
The CE does not, yet the logarithm of the power spectrum appears to achieve a similar goal.

Several more differences between the magnetization and energy CE drops are worth pointing out.
Firstly, for the magnetization CE drop, the hot temperature tail appears higher than its cold counterpart, while for the energy CE drop, it is the opposite. 
Together with the locations of the maxima, this observation indicates that the two CEs might be dual to one another under a map that exchanges high and low temperatures.
Secondly, the increase in maximal CE drop continues to increase for magnetization, while saturating quickly for energy.
The bottom panels of~\ref{fig: CIMs energy} have so much overlap that it is hard to see half the curves.
Under an increase in grid size $L\mapsto(1+\lambda) L$, with $\lambda\ll 1$, the maximum CE changes (to first order in $\lambda$) $\text{CE}_{\max}\mapsto \text{CE}_{\max}+\lambda d$.
For the drop to saturate CE too must change as $\text{CE}\mapsto \text{CE}+\lambda d$, implying that the energy CE is an extensive quantity.
Lack of saturation, as in the magnetic CE, implies the measure is somewhere in between extensive and intensive.

Meanwhile, CC scales well with the total number of spins. 
The top panels of figures~\ref{fig: CIMs magnetization}\& \ref{fig: CIMs energy} show CC per spin for magnetic 
 and energy fluctuations, respectively.
Though both reveal minima near the critical temperature, their locations, at fixed $L$, are most similar to the locations of the CE drop maxima for the magnetization CCs. 
Our simulations confirm that for 2D, the fractional difference between CE and CC extrema is less than 0.5\%, while for 3D, it remains under 1\%.

At high temperatures, magnetic CC density asymptotes toward $\sim 0.5$, particularly evident in 3D. 
In 2D, the trend is less clear, but theoretical considerations suggest an asymptotic value of $(e\ln 2)^{-1} \approx 0.5307$. 
Interestingly, at low temperatures, a secondary CC maximum emerges, which never exceeds $\sim 0.5$ across all lattice sizes.
As with the duality in the CE drops between magnetization and energy, the energy CC density in 3D behaves analogously to the magnetization CC density, with high-low temperatures reversed.
In 2D, a maximum above $T_c$ is not present in the energy CC density like it is below $T_c$ in the magnetization CC density.
In all cases, however, the CC density appears bounded from above in a manner independent of lattice size.
In fact, we now argue for why the value can never exceed the peculiar asymptotic limit mentioned earlier.

How the power spectrum is treated is the key difference between the definitions of our two CIMs.
For CE, equation \ref{eq: CE1}, the spectrum is normalized, while for CC, equation \ref{eq: CC1}, it is scaled to its maximal value.
Using the modal fraction definition (Eq.~\eqref{eq: modal fraction}), we derive:
\begin{align}\label{eq: CIM relationship}
    2^{-I_{*}}\text{CC} = \text{CE} - I_*,
\end{align}
where $I_* \equiv -\log_2 \max_{k'}\mathcal P(k')$ represents the point information, or \textit{surprise}, associated with the dominant mode (see Appendix~\ref{app: 1}).
Equation~\ref{eq: CIM relationship} represents a model dependent relationship between CC and CE since $I_*$ will be determined by the system being examined.
Nonetheless, with a smidgeon of work we can push out this model dependence to derive a model independent bound.

First, since $I_* \leq \log_2 N$, introduce the nat (not bit) based $\mathcal{I} = \ln N - I_{*} \ln 2\ge 0$.
Given the convenience of the scaling used in the figures, define both  $x = \ln N - \text{CE} \ln 2$ and $y = (\text{CC} \ln 2)/N$.
These in hand, eq.~\eqref{eq: CIM relationship} is transformed to (details in Appendix~\ref{app: 2}):
\begin{align}\label{eq: CIM relationship 2}
    y = -e^{-\mathcal{I}}x + \mathcal{I}e^{-\mathcal{I}}.
\end{align}
If $y\rightarrow 0$ (CC$\rightarrow 0$), then either $\mathcal I\rightarrow\infty$ or $x\rightarrow \mathcal I$.
The first case is the thermodynamic limit, $N\rightarrow\infty$, with the maximum power mode scaling slower than $N^{-1}$.
Of course, if it were scaling at $N^{-1}$, then $\mathcal I$ would have to be constant, and we'd find ourself in the second case mentioned, $x=\mathcal I$, or CE$=I_*$.
This is a far more difficult constraint, with particular solutions corresponding to uniform power distributions across subsets of modes and zero power in the rest, which trivially result in a vanishing CC.
These include both the case of uniform power across all modes, and all the power in a single mode, neatly unifying the two limits.
The more general solutions interpolate between these, leading to large families of vanishing complexity.
Though the particular solutions lack complexity, the general solutions are by no means simple.
This rather unexpected result throws into question what the meaning of complexity is for the measure.
Future work should explore the interpretation of simplicity and complexity revealed here.

For nonvanishing CC, CE must be greater than $I_*$, and since the first term in eq.~\eqref{eq: CIM relationship 2} is negative definite, the $\mathcal I e^{-\mathcal I}$ term gives a model dependent upper bound on the complexity.
The model independent upper bound follows directly from there. 
$\mathcal{I}e^{-\mathcal{I}}$ maximizes at $\mathcal{I} = 1$, giving:
\begin{align}\label{eq: CC bound}
    \frac{\text{CC}}{L^d} < \frac{1}{e \ln 2} \approx 0.5307 \frac{\text{bits}}{\text{node}},
\end{align}
Saturating this bound can only occur if $x\rightarrow 0$, and $\mathcal I\rightarrow 1$.
The first of these implies that CE is approaching its maximum value, where the power in each mode is $N^{-1}$.
The second implies that the power in the maximal mode is $eN^{-1}$, in contradiction.
The bound is a soft one which cannot be saturated.

\section{Discussion} \label{sec: V}

This study has systematically examined the behavior of Configurational Information Measures (CIMs) in the Ising model, revealing that both Configurational Entropy (CE) and Configurational Complexity (CC) exhibit distinctive signatures near the critical temperature. 
Our results establish CIMs as analytical tools for understanding phase transitions, offering insight into the redistribution of fluctuation power and the emergence of structured behavior in critical systems.

The observed behavior of magnetic CE aligns with previous studies of Ginzburg-Landau models, confirming that CE effectively tracks phase transitions by quantifying spectral reorganization. 
Our extension to include energy CE and both magnetization and energy CC has revealed several novel insights. 
First, our results point to a duality between magnetization and energy CIMs under temperature inversion across $T_c$. 
Second, our analysis suggests the energy CE is extensive while the magnetization CE exhibits more complex scaling. 
These observations raise important questions about the universal properties of CIMs across different systems. 
Extending this analysis to other universality classes would help distinguish which features are specific to the Ising model versus inherent to CIMs generally.
Perhaps the most significant theoretical contribution of this work is the derivation of both a model-dependent CE-CC relationship and a model-independent upper bound on CC. 
To our knowledge, these represent the first general analytical results establishing fundamental limits on configurational complexity measures. 
The bound $\text{CC} < L^d/(e \ln 2)$ bits/node reveals an intrinsic volume-based limit on complexity independent of the underlying model, suggesting a universal constraint on information-theoretic measures of structural complexity.
Another intriguing finding is the observation of maxima in both energy and magnetization CC density. 
Previous studies have interpreted such maxima in the context of the stability conjecture, correlating them with the onset of instability, while CC minima have been associated with configurations near stability. 
However, our results complicate this interpretation, as the placement of extrema differs depending on whether energy or magnetization CC is examined. 
The proposed duality between CIMs may help reconcile which construction most appropriately aligns with the stability conjecture.

The broader applicability of CIMs extends beyond the Ising model to complex systems lacking clearly defined order parameters. 
Many quantum phase transitions, for instance, defy characterization through conventional symmetry-breaking paradigms, instead relying on entanglement-based measures to identify universal behavior~\cite{PhysRevX.11.041014}. 
CIMs offer an alternative approach by quantifying the structural complexity of fluctuations, potentially revealing phase transitions in systems where the nature of order remains ambiguous. 
Recent applications to lump-like solutions in scalar field models further demonstrate their utility in non-standard phase transitions~\cite{arxiv:2412.00352}.

The apparent contradictions in our findings regarding the meaning of complexity highlight the need for further theoretical development. 
The model-dependent relationship between CE and CC suggests that while these measures capture important aspects of structural organization, their interpretation requires careful consideration of the specific system under study. 
Our analysis reveals families of configurations with vanishing complexity yet non-trivial structure, challenging conventional notions of what "complexity" means in information-theoretic contexts.

In conclusion, our results strongly support the use of CIMs in analyzing complex many-body systems and studying phase transitions and criticality, while simultaneously calling for caution in their interpretation. 
The analytical bounds and relationships established here provide a foundation for more rigorous applications of these measures across diverse physical systems. 
We hope this work serves as both an accessible introduction to CIMs and a catalyst for deeper exploration of why they behave as they do.

\section{Acknowledgement}  
The authors thank the Center for Integrated Research Computing (CIRC) at the University of Rochester for providing computational resources and technical support. 
This project was partly made possible through the support of Grant 62417 from the John Templeton Foundation.
The opinions expressed in this publication are those of the authors and do not necessarily reflect the views of the John Templeton Foundation.

\appendix

\section{Derivation of CC-CE Relationship}\label{app: 1}

The configurational information measures (CIMs), configurational entropy (CE) and configurational complexity (CC), are formally defined as:
\begin{align}\label{eq: A1 1}
    \text{CE} &= -\sum_k  \frac{ \mathcal P(k)}{Z}\log_2\frac{\mathcal P(k)}{Z}, \\ \label{eq: A1 2}
    \text{CC} &= -\sum_k f(k)\log_2 f(k),
\end{align}
where the modal fraction $f(k)$ scales the power spectrum $\mathcal P(k)$ by its maximum value:
\begin{align}\label{eq: A1 3}
    f(k) = \frac{\mathcal P (k)}{\max_{k'}\mathcal P(k')}.
\end{align}
We have assumed that $\mathcal P(k)=|\tilde\phi(k)|^2$ is not normalized to explicitly show how the partition function $Z=\sum_k\mathcal P(k)$ enters in.
Lastly define the \textit{point surprise} associated with the most probable mode as:
\begin{align}
    I_*=-\log_2\frac{\max_k\mathcal P(k)}{Z},
\end{align}
so that the relationship between the normalized power spectrum and the modal fraction is $\mathcal P(k)/Z = 2^{-I_*}f(k)$.
Then, directly from the above definitions
\begin{align}
    \text{CE}&=-2^{-I_*}\sum_k f(k)\log 2^{-I_*} f(k)\nonumber\\
    &=2^{-I_*}\text{CC}+I_* 2^{-I_*}\sum_k f(k)\nonumber\\
    &=2^{-I_*}\text{CC}+I_*,
\end{align}
where in the last line we have used $\sum_kf(k)=2^{I_*}\sum_k\mathcal P(k)/Z = 2^{I_*}$.
Moving the $I_*$ to the LHS establishes equation \eqref{eq: CIM relationship}.

This relationship establishes a direct scaling connection between CE and CC, incorporating the influence of the most dominant spectral mode. 
The presence of $I_*$ indicates that CC is fundamentally tied to how power is distributed across spectral modes, modulating CE based on how concentrated or dispersed the spectral energy is.
Since this will depend on the particulars of the Hamiltonian being examined, the relationship is model dependent.

\section{Reduced Variables}\label{app: 2}

To simplify the CC-CE relationship and analyze its scaling properties, we introduce the following reduced variables:
\begin{align}
    \mathcal{I} &= \ln N - I_* \ln 2,\\
    x &= \ln N - \text{CE} \ln 2,\\
    y &= \frac{\text{CC}}{N} \ln 2.
\end{align}

The first equation represents the effective information content of the system after accounting for the dominance of a particular mode, while the second and third define CE and CC in terms of rescaled variables that incorporate system size.

By subtracting the first two equations, we obtain:
\begin{align}\label{eq: app2}
    \text{CE} - I_* = \frac{\mathcal{I} - x}{\ln 2}.
\end{align}
Rewriting the third equation by inverting its definition gives:
\begin{align}\label{eq: app3}
    \text{CC} &= \frac{y e^{\ln N}}{\ln 2} = \frac{y e^{\mathcal{I}} 2^{I_*}}{\ln 2}.
\end{align}

Substituting Eqs.~\eqref{eq: app2} and \eqref{eq: app3} into Eq.~\eqref{eq: CIM relationship} leads to the reduced form of the CIM relationship:
\begin{align}
    y e^{\mathcal{I}} = \mathcal{I} - x.
\end{align}
This expression explicitly shows how the configurational measures interact via an exponential scaling factor, emphasizing the role of information concentration in modulating complexity. 
\end{document}